\documentclass[showpacs,twocolumn,aps]{revtex4}
\usepackage{graphicx}
\usepackage{amsmath,subfigure}
\usepackage{verbatim}
\usepackage{amssymb}
\newcommand{\Avg}[1]{\big\langle{#1}\big\rangle}

\newcommand{\beas}{\begin{eqnarray*}}
\newcommand{\eeas}{\end{eqnarray*}}
\newcommand{\cavg}[1]{\left\langle\!\left\langle{#1}
        \right\rangle\!\right\rangle}
\renewcommand{\l}{\left}
\renewcommand{\r}{\right}

\newcommand{\Tr}{{\rm Tr}}
\newcommand{\erf}{{\rm erf}}
\newcommand{\erfc}{{\rm erfc}}
\newcommand{\disavg}[1]{\left[{#1}\right]_{{\rm dis}}}

\newcommand{\ii}{{\rm i}}
\newcommand{\req}[1]{(\ref{#1})}
\newcommand{\be}{\begin{equation}}
\newcommand{\ee}{\end{equation}}
\newcommand{\bea}{\begin{eqnarray}}
\newcommand{\eea}{\end{eqnarray}}
\newcommand{\dd}{\textrm{d}}
\newcommand{\pr}[1]{\left(#1\right)}
\newcommand{\cro}[1]{\left[#1\right]}
\newcommand{\acc}[1]{\left\{#1\right\}}

\newcommand{\avg}[1]{\langle{#1}\rangle}
\newcommand{\ovl}[1]{\overline{#1}}
\newcommand{\BE}{\begin{eqnarray}}
\newcommand{\EE}{\end{eqnarray}}
\newcommand{\BEn}{\begin{eqnarray*}}
\newcommand{\EEn}{\end{eqnarray*}}
\newcommand{\barr}{\begin{array}}
\newcommand{\earr}{\end{array}}

\newcommand{\bit}{\begin{itemize}}      
\newcommand{\eit}{\end{itemize}}
\newcommand{\bc}{\begin{center}}
\newcommand{\ec}{\end{center}}
\newcommand{\ben}{\begin{enumerate}}    
\newcommand{\een}{\end{enumerate}}

\newcommand{\eps}{\epsilon}

 \newcommand{\e}{\mbox{e}}

\begin{document}

\title{Minority games with finite score memory} \author{Damin
Challet$^1$, Andrea De Martino $^2$, Matteo Marsili $^3$ and Isaac Perez
Castillo$^4$} \affiliation{$^1$Nomura Centre for Quantitative Finance,
Mathematical Institute, Oxford University, 24--29 St Giles', Oxford
OX1 3LB, United Kingdom\\$^2$INFM-SMC and Dipartimento di Fisica,
Universit\`a di Roma ``La Sapienza'', P.le A. Moro 2, 00185 Roma,
Italy\\$^3$The Abdus Salam International Centre for Theoretical
Physics, Strada Costiera 11, 34100 Trieste, Italy\\$^4$Instituut voor
Theoretische Fysica, Katholieke Universiteit Leuven, B-3001 Leuven,
Belgium}

%\date{~}

%%%%%%%%%%%%%%%%%%%%%%%%%%%%%%%%%%%%%%%%%%%%%%%%%%%%%%%%%%%%%%%%%%%%%%%%%%%%%

\begin{abstract}
We analyze grand-canonical minority games with infinite and finite
score memory and different updating timescales (from `on-line' games
to `batch' games) in detail with various complementary methods, both
analytical and numerical. We focus on the emergence of `stylized
facts' and on the production of exploitable information, as well as on
the dynamic behaviour of the models. We find that with finite score memory no agent can be frozen, and that all the current analytical methods 
fail to provide satisfactory explanation of the observed behaviours.
\end{abstract}

\maketitle 

\section{Introduction}

Few realistic agent-based models of financial markets can be
understood in depth. Among these, minority games (MGs)
\cite{CZ97,Book} are perhaps the most studied at a fundamental
physical level, especially through a systematic use of spin-glass
techniques \cite{CMZe99,dm,CoolenBatch,hd,CoolenOnline}. The inclusion
of many important aspects of market dynamics in the standard MG setup,
however, often leads to considerable technical difficulties and raises
new challenges for statistical mechanics. A particularly important
modification concerns the memory of agents. In the original game, the
learning dynamics on which traders base their strategic decisions is
such that they remember all their past payoffs irrespective of how far
in time they occurred. It is however reasonable to think that real
traders tend to base their choices only on the most recent
events. This rather natural extension was originally introduced in
\cite{MMRZ} for a model in which agents play the MG strategically. For
the purpose of modeling financial markets, the relevant situation is
instead that of price-taking, or na\"\i ve, agents. In this case, it
has been argued that finite score memory gives rise to a surprisingly
rich dynamical phenomenology \cite{CDM04}. In this paper, we will
analyze such models in detail.

There are of course several ways to introduce a finite memory in the
MG. A conceptually simple one is to fix a time window $M$ (the `score
memory') during which agents keep exact track of their scores
\cite{J00,JCrash}. The main advantage is that the game becomes
Markovian of order $M$. This situation can be handled numerically for
reasonably small $M$. Here, we are interested in the case in which $M$
is of the order of the number of traders $N$, which is supposed to be
very large (ultimately, the limit $N\to\infty$ will be considered).
For the sake of simplicity, we choose to implement the situation in
which scores are exponentially damped in time, which requires only a
minor modification of the original equations. Furthermore, we shall
focus on the grand-canonical MG (GCMG) \cite{CM03}, which is known to
produce market-like fluctuation phenomena and whose properties have
been shown to be extremely sensitive to the introduction of a finite
memory \cite{CDM04}. Both the `on-line' and `batch' versions of the
model will be addressed. The two situations differ by the timescales
over which agents update their status: in the former, the updating
takes place at every time step; in the latter, it occurs roughly once
every $P$ time steps, with $P=\mathcal{O}(N)$. In particular,
following \cite{TobiasInterpol}, we use a parameter that interpolates
between the `on-line' and `batch' models and study how the resulting
fluctuation phenomena are affected by changes of updating timescales.

We shall proceed by defining the models (Sec. II) and exploring, via
computer experiments, their behaviour (Sec. III). We will focus
especially on the emergence and parameter-dependence of `stylized
facts', peculiar statistical regularities that are empirically
observed in financial markets. In Sections IV and V we will
characterize analytically the stationary states of the model with
infinite and finite memory, respectively. We use static
replica-based minimization techniques in order to compute the
properties of the on-line game with infinite score memory. For the batch game, one can
resort to dynamical mean-field theory. The resulting theory is exactly
solvable in the case of infinite memory. For finite-memory games, we
have to resort a Monte Carlo scheme, known as Eissfeller-Opper method
\cite{EO}, to extract the steady state from the dynamical mean-field
equations. Finally, we formulate our concluding remarks in Sec. VI. For
the sake of completeness, we add an Appendix in which the effects of
finite score memory in the standard MG are discussed.

\section{Definition of the model}

In a minority game, at each time step $t$, $N$ traders are faced with
two choices: to buy or to sell; those who happen to be in the minority
win. The game is repeated, and the traders' actions are determined by
a simple re-inforcement learning. Each agent $i$ has his own fixed
trading strategy $\boldsymbol{a}_i=\{a_i^\mu\}$ that specifies an
action $a_i^\mu\in\{-1,1\}$ for each of the $\mu=1,\ldots,P$ states of
the world. Each component of every strategy is randomly drawn from
$\{-1,1\}$ with uniform probability before the beginning of the
game. The adaptation abilities of agent $i$ are limited to choosing
whether to participate or to withdraw from the market, denoted
respectively by $n_i(t)=1$ and $n_i(t)=0$. At time $t$, the state of
the world $\mu(t)$ is drawn equiprobably from $\{1,\ldots,P\}$. Agent
$i$ sets his $n_i(t)$ according to the sign of his strategy score,
denoted by $U_i(t)$. In particular, he plays $n_i(t)a_i^{\mu(t)}$
where $n_i(t)=\Theta[U_i(t)]$ and $\Theta(x)$ is the Heaviside step
function. The total excess demand $A(t)$ at time $t$, namely the
numerical difference between buyers and sellers, is
\begin{equation}
A(t)=\sum_{i=1}^N n_i(t)a_i^{\mu(t)}.
\end{equation}
All the agents then update their score according to
\begin{equation}\label{on-line}
y_i(t+1)=\l(1-\frac{\lambda_i}{P}\r)y_i(t)-
\frac{1}{P}a_i^{\mu(t)}A(t)-\frac{\epsilon_i}{P}
\end{equation}
with $\lambda_i\geq 0$ a constant. The $(1-\lambda_i/P)$ term is
responsible for finite score memory: if $\lambda_i=0$, the agent has
infinite score memory, as usually assumed in MGs, while for
$\lambda_i>0$ he has an exponentially damped memory of his past
scores.  More precisely, the number of time steps needed to forget a
fraction $f_i$ of his/her payoff is $\ln(1-f_i)/\ln(1-\lambda_i/P)$;
for $\lambda_i/P\ll 1$, it is proportional to $P/\lambda_i$.  The
payoff $~-a_i^{\mu(t)}A(t)$ is that of a Minority Game. The last term
$\eps_i$ sets a benchmark that agent $i$ has to beat in order to
participate in the market. For instance, $\eps_i$ can be thought of as
the interest of a risk-free account (see \cite{CM03,CMZ01} for more
details).

Although our analysis can be extended easily to a more heterogeneous
case, we assume, for the sake of simplicity, that $\lambda_i=\lambda$
and that there are two groups of agents: those with $\eps_i=-\infty$,
referred to as `producers', who always take part in the market, and
the rest, who have $\eps_i=\eps$ finite and are called `speculators'
\cite{CMZ00}. Traders with $\epsilon>0$ (resp. $\epsilon<0$) are
risk-averse (resp. risk-prone), i.e., they have an incentive to stay
out of (resp. enter) the market. We denote by $N_s$ and $N_p$ the
number of speculators and producers, respectively.  The producers,
being deterministic, inject information into the market, that the
speculators try to exploit. This setup, which defines an ecology of
market participants, has been introduced in \cite{CMZ01} as the
simplest tractable interacting agents model able to reproduce the
`stylized facts' of financial markets \cite{CM03}. In the statistical
mechanics approach, one is interested in the limit of large systems,
in which $P,N_s,N_p\to\infty$ keeping the reduced number of agents
$n_s=\lim_{P\to\infty}N_s/P$ and $n_p=\lim_{P\to\infty}N_p/P$
fixed. 

Up to now, agents update their variables $n_i(t)$ at each time
step. It is natural in financial markets to assume that traders prefer
not to change their strategy every time step in order to avoid
over-reacting, and also because estimating the performance of a
strategy needs some time. This can be approximated in our model by
allowing agents to change their variable $n_i$ every $T$ time steps
\cite{TobiasInterpol}. For the sake of simplicity, we assume that
agents perform the updates synchronously. If $T\gg P$, the score
update between $t$ and $t+T$ is essentially an average of score
increases over all the states of the world. In this limit, defining
$t'$ as $t/T$, one can rewrite \req{on-line} as
\begin{equation}\label{batch}
y_i(t'+1)=(1-\lambda_i)y_i(t')-\sum_{j=1}^N J_{ij}
n_j(t')-\alpha\epsilon_i
\end{equation}
with $\alpha=P/N=(n_s+n_p)^{-1}$ and quenched random couplings
$J_{ij}$ given by
\begin{equation}\label{couplings}
J_{ij}=\frac{1}{N}\sum_{\mu=1}^P a_i^\mu a_j^\mu
\end{equation}
Because this dynamics is equivalent to enumerating all
$\mu\in\{1,\ldots,P\}$, we assumed that all the states of the world
always occur between $t'$ and $t'+1$.  The neural network literature
refers to models with $T=1$ as `on-line', whereas models such as
\req{batch} are called `batch'. The parameter $T$ allows the
interpolation between the former and the latter. 

Batch minority games were introduced in \cite{Oxf2} and solved later
exactly with generating functionals \cite{CoolenBatch}, an exact
dynamical method. The stationary states of (\ref{on-line}) and
(\ref{batch}) are, strictly speaking, different even in the
thermodynamic limit. This is because agents in batch games have a
longer auto-correlation: for instance, denoting $\phi_i=\avg{n_i}$ the
time average of $n_i(t)$ in the stationary state, one has $\avg{n_i
n_j}=\phi_i\phi_j$ in on-line games \cite{CMZe99,CoolenOnline}, but
not in batch games \cite{CoolenOnline}.

\section{Stylized facts}

The connection between minority game's outcome $A$ and real prices
comes from relating $A$ to the `excess demand', and linking the
evolution of the price $p(t)$ to it via
\begin{equation}
\log p(t+1)=\log p(t)+\frac{A(t)}{L}
\end{equation}
where $L$ is a constant, the `liquidity', that will be hereafter fixed
to $1$ \cite{Farmer,J99}. While the original MG revealed insightful
relationships between price fluctuations and predictability, it fails
to reproduce the empirically observed market-like behaviour, in
particular the so-called `stylized facts'. As a consequence, GCMGs
were introduced to better mimic market dynamics.  They are {\em able}
to produce stylized facts such as fat-tailed price return
distributions $P(A)\propto A^{-\beta}$ with $\beta\simeq 3.5$ and
volatility clustering $\avg{A(t)^2A(t+\tau)^2}\propto\tau^{-\gamma}$
with $\gamma\simeq 0.3$; these exponents are close to those measured
in real markets. They are however unable to reproduce well documented
over-diffusive price behavior \cite{Daco,BouchaudPotters}, because the
MG induces a mean-reverting process.

The presence of stylized facts in the GCMG with $\lambda=0$ was linked
to a too small signal-to-noise ratio, suggesting that marginal
efficiency is a necessary condition for the existence of stylized
facts \cite{CM03}. In other words, in infinite systems, stylized facts
only occur at a the phase transition, whereas in finite systems,
critical-like phenomenology is observed in a critical window around
the critical point, which shrinks as the system size increases (see
\cite{CM03} for more details and for a way of keeping alive stylized
facts in infinite systems). Therefore, the stylized facts of this
model cannot be studied by the methods of statistical mechanics.

\begin{figure}
\centerline{\includegraphics*[width=8.5cm]{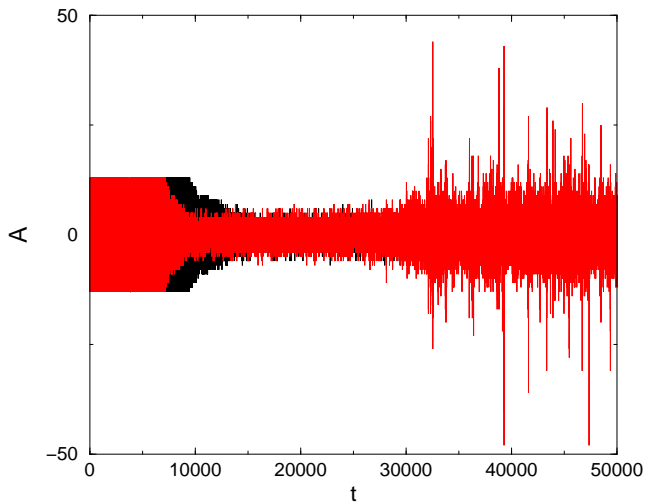}}
\caption{Return as a function of time for the same realisation of a
GCMG with biased initial strategy scores and $\lambda=0$ (black
lines), and $\lambda=0.003$ (red lines). System set in the critical
window, $P=25$, $n_s=40$, $\eps=0.01$}\label{fig:At}
\end{figure}

While the {\em ability} of the GCMG to produce stylized facts was
emphasised in previous work, Ref. \cite{CDM04} pointed out two
delicate problems of GCMG regarding stylized facts. The first one is
the dependence of stylized facts on initial conditions: assume that
one given realization of the game produces stylized facts; changing
slightly the initial conditions $y_i(0)$ is then enough to destroy
them, leading to scenarios with only Gaussian price changes. The
second problem is the following: for a given set of parameters, one
realization of the game may produce stylized facts but another
not. Both problems are due to the coincidence of fixed disorder in the
strategies and infinite score keeping. This ultimately motivates the
introduction of finite score memory. Fig. \ref{fig:At} compares $A(t)$
for the same game with $y_i(0)\ne 0$, once with $\lambda=0$ and once
with $\lambda>0$. In the latter case, fluctuations have a
characteristic pattern as a function of time: they first decrease to a
very small value, stay at this level for a time interval of order
$1/\lambda$ and then increase, producing fat-tailed price returns. The
same kind of volatility behavior occurs in the original MG (see
Appendix \ref{appendix:MG}).

\begin{figure}
\centerline{\includegraphics*[width=8.5cm]{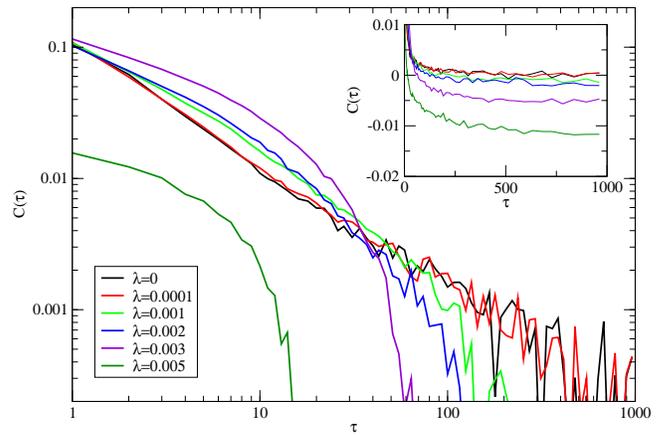}}
\caption{Absolute-valued price return auto-correlation function for
increasing $\lambda$ for a given realisation of an on-line game;
inset: same data in double-linear scale ($10^7$ iterations per run,
after $200P$ iterations; $P=20$, $n_s=20$, $n_p=1$,
$\eps=0.01$)}\label{P(A)corr}
\end{figure}

\begin{figure}[b]
\centerline{\includegraphics*[width=8.5cm]{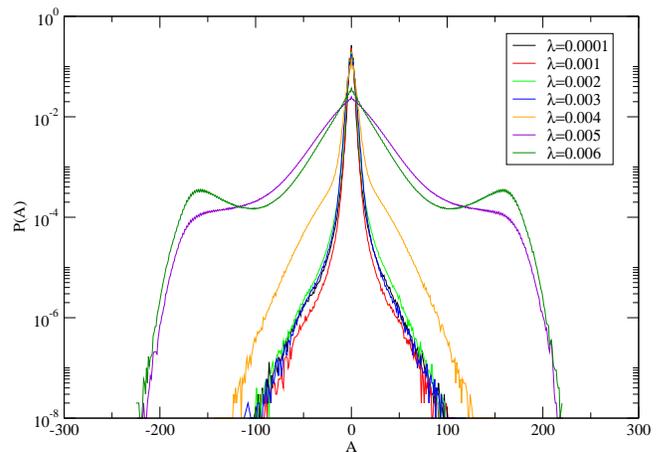}}
\caption{Average price return distribution function of on-line games
for increasing $\lambda$. Average of 100 samples on $10^6$ iterations
per run, after $200P$ iterations; $P=20$, $n_s=20$, $n_p=1$,
$\eps=0.01$.}\label{P(A)online}
\end{figure}

However, finite score memory is a double-edged sword. First of all,
certain empirical stylized facts, in particular `volatility
clustering', suggest the presence of long-memory effects in
markets. If one defines volatility clustering via the requirement
\begin{equation}
C(\tau)\equiv\frac{\avg{|A(t)||A(t+\tau)|}}{\avg{A(t)^2}}\simeq
\tau^{-\gamma}
\end{equation}
it is easily understood that any positive value of $\lambda$ destroys
this power-law dependence, because of the cut-off that it imposes at
times of the order of $P/|\ln(1-\lambda)|$. Fig. \ref{P(A)corr}, by
displaying $C(\tau)$ for $\tau\le1000$, is not able to show this
effect for $\lambda<0.001$; the cut-off is clear for larger
$\lambda$. Not only $C(\tau)$ is cut off, but the loss of memory
induces a negative $C(\tau)$ at large $\tau$ (see inset).

\begin{figure}
\centerline{\includegraphics*[width=8.5cm]{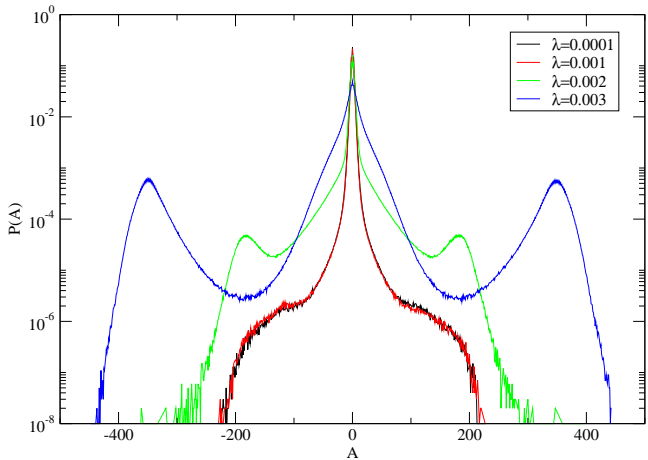}}
\caption{Average price return distribution function of on-line games
for increasing $\lambda$. Average of 100 samples on $10^6$ iterations
per run, after $200P$ iterations; $P=20$, $n_s=40$, $n_p=1$,
$\eps=0.01$.}\label{P(A)online2}
\end{figure}

The other potentially nefarious effect of finite score memory is
threaten the power-law tails: GCMGs produce power-law tailed $A$
because of a volatility feed-back \cite{CM03}. This feed-back needs
some time to establish, hence if the memory length associated to
$\lambda$ is smaller than this time, power-law tails should
disappear. This is indeed the case (see Figs \ref{P(A)online} and
\ref{P(A)online2}), in a queer way: the central part of $P(A)$ is
exponential, but the size of the support of $P(A)$ actually increases
because of the appearance of two peaks. Note that maximum value
$\lambda^*$ of $\lambda$ for which stylized facts are preserved
depends on the system's parameter, as shown by these two figures.

We conclude that, in practice, a sufficiently small
$\lambda\ge\lambda^*\simeq 0.001$ preserves the salient stylized
facts: the noise of $C(\tau)$ for $\tau\ge 100$ in GCMG or in
financial market data is such that the values of $\lambda$ that
preserve power-law tails of $P(A)$ do so for $C(1)$. Therefore, the
introduction of finite score memory does not affect significantly the
market-like phenomenology produced by GCMG. The value of $\lambda^*$
obtained has to be contrasted with the small typical time-window used
e.g. in \cite{J00,JCrash}.

\begin{figure}
\centerline{\includegraphics*[width=8.5cm]{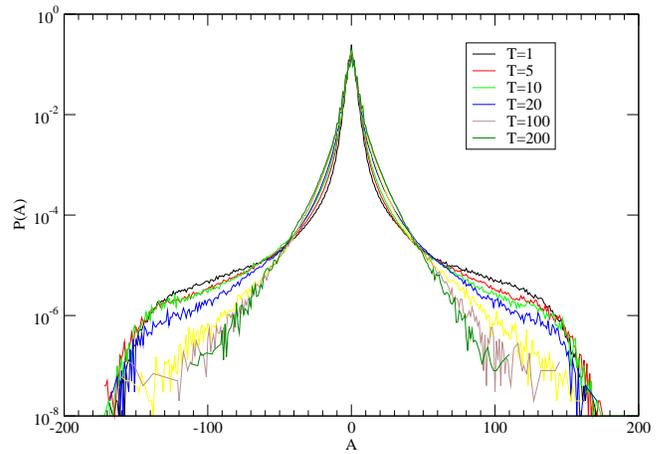}}
\caption{Average price return distribution function for increasing
$T$. Average of 100 samples on 1,000,000 iterations per run, after
$200P$ iterations; $P=20$, $n_s=40$, $n_p=1$,
$\eps=0.01$.}\label{P(A)T}
\end{figure}

\begin{figure}
\centerline{\includegraphics*[width=0.35\textwidth,angle=270]{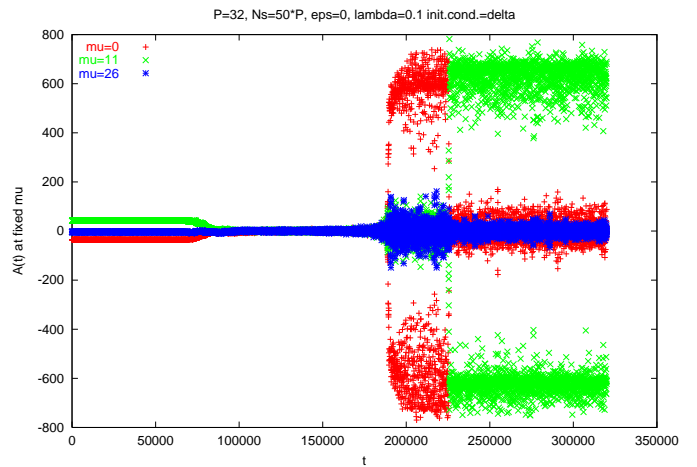}}
\caption{Price return as a function of time for various patterns
$\mu$. A symbol is plotted only when
$\mu(t)=\mu$.}\label{fig:finmempattern}
\end{figure}

Let us now investigate the effect of updating the $n_i(t)$ every $T$
time steps (see Fig. \ref{P(A)T}). Increasing $T$, one interpolates
between on-line games ($T=1$) and batch games ($T\gg P$). One clearly
sees that this destroys large price changes: while $P(A)$ has
power-law tails when $T=1$, it is gradually transformed into an
exponential distribution for $T\gg P$. This is because large price
changes is associated to a given pattern $\mu$ \cite{CDM04,JCrash}
that varies as a function of time, as illustrated by
Fig. \ref{fig:finmempattern}. However, it should be noted that
$T\simeq P$ gives rise to a cleaner power-law.

\begin{figure}[t]
\centerline{\includegraphics*[width=8.5cm]{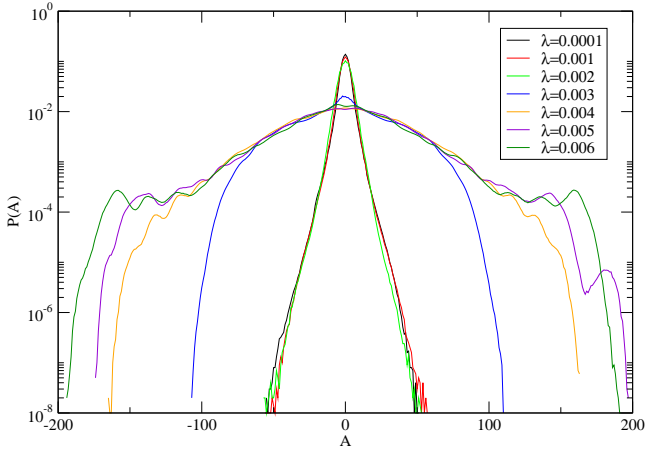}}
\caption{Average price return distribution function of batch games for
increasing $\lambda$. Average of 100 samples on 1,000,000 batch iterations
per run, after $200P$ batch iterations; $P=20$, $n_s=20$, $n_p=1$,
$\eps=0.01$.}\label{P(A)batch}
\end{figure}

Batch games display the same kind of transition when $\lambda$ is
increased.  Fig.  \ref{P(A)batch} shows that the largest value of
$\lambda$ such that $P_\lambda(A)\simeq P_0(A)$ is smaller and around
$0.002$ for $n_s=20$, $P=20$. Interestingly, the distribution of price
changes $P(A)$ is more or less stable with respect to $\lambda$ as
long as $\lambda<0.004$ in this figure. This is an important condition
for the use of finite memory in these models.

\section{Stationary state with $\lambda=0$}

This section characterizes the steady state of both the on-line
(Eq. \req{on-line}) and batch GCMG (Eq. \req{batch}) for
$\lambda=0$. The relevant macroscopic observables are as usual the
`volatility' $\sigma^2$ and the `predictability' $H$, given
respectively by
\begin{equation}
\sigma^2=\frac{\avg{A^2}}{P}~~~~~~~~~~~
H=\frac{\ovl{\avg{A|\mu}^2}}{P}
\end{equation}
where $\avg{\cdots}$ and $\avg{\cdots|\mu}$ denote time averages in
the stationary state, the latter conditioned on the occurrence of the
piece of information $\mu$, and the over-line denotes an average over
information patterns. In on-line games, these two quantites are linked
by \be \sigma^2=H+\phi-G \ee where $G=\sum_{i=1}^{N_s}\phi_i^2/P$,
$\phi_i=\avg{n_i}$ denoting the probability that speculator $i$ joins
the market in the steady state.  The normalizing $P$ factors have been
introduced in order to ensure that all quantities remain finite when
$P$, $N_s$, $N_p\to\infty$. In addition to these, an important role is
played by the relative number of active speculators
\begin{equation}
\phi=\frac{1}{N_s}\sum_{i=1}^{N_s}\phi_i
\end{equation}
as well as by the number of active speculators per pattern,
i.e. $n_{\text{act}}=n_s\phi$.

\subsection{On-line GCMG: static approach}
\label{Stat:on-line}

Partial results from replica-based calculus are reported in
Ref. \cite{CM03}, and are based on the existence of a global quantity
$H_\eps$ that is minimized by the dynamics. Finding $H_\eps$ relies on
the prescription given in Ref. \cite{MC01}: first derive the
continuous-time stochastic differential version of Eq. \req{on-line},
which reads \be\label{on-line:conttime} \dot
y_i(\tau)=-\ovl{a_i\avg{n_iA}}-\eps+\eta_i(\tau) \ee where $\tau=t/P$
is the intrisic time of the GCMG, $\eta_i(t)$ is a zero-average
Gaussian noise with $\avg{\eta_i(\tau)\eta_j(\tau')}=\frac{1}{N}
\ovl{a_i a_j \avg{A^2}_y}\delta(\tau-\tau')$, and the over-line
denotes an average over information patterns: $\ovl{\cdots}=(1/P)
\sum_{\mu=1}^P\cdots$. The deterministic term of this equation can be
interpreted as the gradient of $H_\eps$, so that
$H_\eps=H_0+2\eps\sum_{i=1}^{N_s}\phi_i$. Stationary states correspond
then to the minima of $H_\eps$. When the stationary state is unique,
its equilibrium properties are entirely determined by $H_\eps$;
otherwise, one has to supplement $H_\eps$ with a self-consistent
equation for computing correlations \cite{MC01}. Regarding $H_\eps$ as
a cost function, one may compute the minima of $H_\eps$ from the
partition function $Z(\beta)=\Tr ~\e^{-\beta H_\eps}$. The typical
properties of the minimum of $H_\eps$, i.e. of the $\beta\to\infty$
limit of free energy, require the evaluation of the quenched-disorder
average $\disavg{\log Z}$, which is performed via the replica trick
$\disavg{\log Z}=\lim_{n\to0}\log\disavg{\avg{Z^n}}/n$.

Reference \cite{CM03} reported plots of the exact solution. Here we
give the final results of the calculus only, as the latter is standard
(see however \cite{CMZ00} for more details). With replica symmetric
ansatz, the free energy $f=(1/\beta\partial) \ln Z/\partial \beta$,
which corresponds to $H_\eps$ in the limit $\beta\to\infty$, is given
by \bea f(g,r)&=&\frac{\alpha}{2\beta}
\log\left[1+\frac{2\beta(G-g)}{\alpha}\right]+\frac{\rho+g}{1+\chi}\nonumber\\
&+&\frac{\alpha\beta}{2}(RG-rg) -\frac{1}{\beta} \Avg{\log\int_{0}^1
d\pi\; e^{-\beta V_z(\pi)}} \eea where we found it convenient to
define the ``potential'' \be V_z(\pi)=-\frac{\alpha\beta(R-r)}{2}
\pi^2-\sqrt{\alpha r}\,z\,\pi+2\eps\pi \ee so that the last term of
$f$ looks like the free energy of a particle in the interval $[0,1]$
with potential $V_z(\pi)$ where $z$ plays the role of disorder.
$G=\sum_{i=1}^{N_s}\phi_i^2$ is the self-overlap and $g$, $R$, $r$ are
Lagrange multipliers.

The four saddle point equations have exactly the same form as MG
without the 0 strategy: \bea\label{r} \frac{\partial f}{\partial
g}=0~~~ &\Rightarrow &~~~~ r=\frac{4(\rho+g)}{\alpha^2(1+\chi)^2}\\
\frac{\partial f}{\partial G}=0~~~ &\Rightarrow &~~~~
\label{R-r}\beta(R-r)=-\frac{2}{\alpha(1+\chi)}\\
\frac{\partial f}{\partial R}=0~~~  &\Rightarrow &~~~~
G=\avg{\avg{\pi^2}_\pi}_z\\
\frac{\partial f}{\partial r}=0~~~  &\Rightarrow &~~~~
\label{G-g}\beta(G-g)=\frac{\avg{\avg{\pi z}_\pi}_z}{\sqrt{\alpha r}}
\eea

In the limit $\beta\to 0$ we can look for a solution with $g\to G$ and
$r\to R$. It is convenient to define \be\label{chizeta}
\chi=\frac{2\beta(G-g)}{\alpha},~~~\hbox{and}~~~
\zeta=-\sqrt{\frac{\alpha}{r}}\beta(R-r) \ee and to require that they
stay finite in the limit $\beta\to\infty$. The potential can then be
rewritten as \be V_z(\pi)=\sqrt{\alpha
r}\left[\zeta\frac{\pi^2}{2}-\pi\left(z-\frac{2\eps}{\sqrt{\alpha
r}}\right)\right] \ee

The averages are easily evaluated since, in this case, they are
dominated by the minimum of the potential $V_z(\pi)$. Let $K$ be
$\eps(1+\chi)$, the minimum of $V_z(\pi)$ is at $\pi=0$ for $z\le
\zeta\ K$ and at $\pi=+1$ for $z\ge \zeta(1+K)$.  For $\zeta\
K<z<\zeta[1+K]$, the minimum is at $\pi=z/\zeta-K$. With this we find,
\be \avg{\avg{\pi z}}=\frac{1}{2\zeta}\
\acc{\erf[(1+K)\zeta/\sqrt{2}]-\erf(K\zeta/\sqrt{2})} \ee and \bea
\avg{\avg{\pi^2}}&&=G=\frac{1}{\zeta\sqrt{2\pi}}\cro{(K-1)\e^{-(1+K)^2\zeta^2/2}-K\e^{-K^2
z^2/2}}\nonumber\\&&+\frac{1}{2}\pr{{K^2+\frac{1}{\zeta^2}}}\left(\erf[(1+K)\zeta/\sqrt{2}]-\erf(K\zeta/\sqrt{2})\right)\nonumber\\&&+\frac{1}{2}\;\erfc[(1+K)\zeta/\sqrt{2}]
\label{eqQ}
\eea

The fraction of agents who never enter into the market is then
$\phi_0=(1+\erf(\zeta K/\sqrt{2})/2$, and the ones who always
participate is $\phi_1=\erfc[(1+K)\zeta /\sqrt{2}]/2$. The pdf of the
$\pi_i$ is given by \be {\cal P}(\pi)=
\phi_0\delta(\pi)+\phi_1\delta(\pi-1)+ \frac{\zeta}{\sqrt{2
\pi}}e^{-(\zeta(\pi+K))^2/2}
\label{Prob}
\ee And the average number of agents in the market $\phi=\avg{\pi}$
where the average is over ${\cal P}(\pi)$, is \bea
\phi&=&\phi_1+\frac{1}{\zeta\sqrt{2\pi}}\pr{\e^{-K^2\zeta^2/2}-\e^{-\zeta^2(1+K)^2/2}}\nonumber\\&&+\frac{K}{2}(\erf(K\zeta/\sqrt{2})-\erf[(1+K)\zeta/\sqrt{2}])\nonumber
\eea Observing that $\zeta=\sqrt{\alpha/(\rho+G)}$, one finally finds
that $\zeta$ is fixed as a function of $\alpha$ and $\rho$ by the
equation \bea
\frac{\alpha}{\zeta^2}&=&\rho+\frac{1}{\zeta\sqrt{2\pi}}\cro{(K-1)\e^{-(1+K)^2\zeta^2/2}-K\e^{-K^2
z^2/2}}\nonumber\\&&+\frac{1}{2}\pr{{K^2+\frac{1}{\zeta^2}}}\left(\erf[(1+K)\zeta/\sqrt{2}-\erf(K\zeta/\sqrt{2})\right)\nonumber\\&&+\frac{1}{2}\;\erfc[(1+K)\zeta/\sqrt{2}]
\eea which has to be solved numerically. With some more algebra, one
easily finds~: \be
K=\eps\left[1-\frac{\erf[(1+K)\zeta/\sqrt{2}]-\erf(K\zeta/\sqrt{2})}{2\alpha}\right]^{-1}
\label{eqx}
\ee These two last equations form a close non-linear set of equations.

The calculus gives finally $H_\eps=\lim_{\beta\to\infty f}$
\be\label{eqH} H_\eps=\frac{n_p+n_sG(\zeta,K)}{(1+\chi)^2}+2\eps
\phi(K,\zeta) \ee

The stationary is unique if $H_\eps\ne 0$, which is the case as long
as $\eps\ne0$ and for $n_s\le n_s^*(n_p,\eps)$ if $\eps=0$.

The fluctuations are given by \be \sigma^2=\eps^2\
\frac{n_p+n_sG}{K^2}+n_s(\phi-G) \label{25}.  \ee

One can also show that $H_\eps\propto \eps^2$ for small $\eps$:
according to Eq. \req{eqH}, this hold if $K\to0<K_0<\infty$, which can
be seen by expanding Eq. \req{eqx} in powers of $\eps$.

\subsection{Batch GCMG: dynamical approach}

We consider the batch-GCMG dynamics, which we re-cast as
\begin{equation}\label{dyn}
y_i(t+1)=y_i(t)-\sum_{j=1}^N
J_{ij}n_j(t)-\alpha\epsilon_i+h_i(t)
\end{equation}
with
\begin{equation}
\epsilon_i=\begin{cases}\epsilon&\text{for $1\leq i\leq
N_s$}\\-\infty&\text{for $N_s+1\leq i\leq N_s+N_p\equiv N$}
\end{cases}\nonumber
\end{equation}
The subscripts $s$ and $p$ denote speculators and producers,
respectively. The relevant dynamical variable is
$n_i(t)=\Theta[y_i(t)]$ (for producers, $n_i(t)=1$), while the
random couplings $J_{ij}$ are given by (\ref{couplings}),
$J_{ij}=(1/N)\sum_\mu a_i^\mu a_j^\mu$, with $a_i^\mu\in\{-1,1\}$ iid
quenched random variables with uniform probability distribution. For
simplicity, we set
\begin{equation}
\alpha=\frac{P}{N}=\frac{1}{n_s+n_p}
\end{equation}
with $n_s=N_s/P$ and $n_p=N_p/P$. The external sources $h_i(t)$ have
been added in order to probe the system against small
perturbations. In the following, we denote averages over all possible
time-evolutions (paths), i.e. realizations of (\ref{dyn}), by double
brackets $\cavg{\cdots}$.

The standard tool for investigating the dynamics of statistical
systems with quenched disorder is the path-integral method \`a la De
Dominicis, based on the evaluation of the generating functional
\begin{equation}\label{gf}
Z[\boldsymbol{\psi}]=
\disavg{\cavg{\exp\l[-\ii\sum_{i,t}n_i(t)\psi_i(t)\r]}}
\end{equation}
from which disorder-averaged site-dependent correlation functions of
all orders can be derived via such identities as
\begin{gather}
\disavg{\cavg{n_i(t)}}=\ii
\lim_{\boldsymbol{\psi}\to\boldsymbol{0}}\frac{\partial
Z[\boldsymbol{\psi}]}{\partial\psi_i(t)}\label{iden}\\
\disavg{\cavg{n_i(t)
n_j(t')}}=-\lim_{\boldsymbol{\psi}\to\boldsymbol{0}}\frac{\partial^2
Z[\boldsymbol{\psi}]}{\partial\psi_i(t)\partial\psi_j(t')}
\end{gather}
In turn, macroscopic (auto)-correlation and response functions like
\begin{gather}\label{c}
C(t,t')=\lim_{N\to\infty}\frac{1}{N}\sum_{i=1}^N\disavg{\cavg{n_i(t)
n_i(t')}}\\G(t,t')=\lim_{N\to\infty}\frac{1}{N}
\sum_{i=1}^N\frac{\partial\disavg{\cavg{n_i(t)}}}{\partial h_i(t')}
\label{g}
\end{gather}
can in principle be evaluated by simply taking derivatives of $Z$ with
respect to the sources $\{\psi_i,h_i\}$. The calculation of $Z$ in the
limit $N\to\infty$ leads, via a saddle-point integration, to the
identification of a non-Markovian single (`effective') agent process
that provides a complete description of the original Markovian
multi-agent process (\ref{dyn}). This procedure requires a
straightforward variation of the cases dealt with in the existing
literature and we will not report it in detail. For our purposes, it
will suffice to say that the effective dynamics for speculators is
given by
%\begin{widetext}
\begin{multline}\label{eff}
y(t+1)=y(t)-\alpha\sum_{t'\leq t}(I+G)^{-1}(t,t')n(t')\\
-\alpha\epsilon+h(t)+\sqrt{\alpha}z(t)
\end{multline}
%\end{widetext}
with $n(t)=\Theta[y(t)]$, whereas $n(t)=1$ for producers always.
Here, $I$ is the identity matrix, $G$ is the response function
(\ref{g}), and $z(t)$ is a Gaussian noise with zero average and time
correlations
\begin{equation}\label{pop}
\avg{z(t)z(t')}=[(I+G)^{-1}C(I+G^T)^{-1}](t,t')
\end{equation}
with $C$ the correlation function (\ref{c}). Notice that the coupling
between the two groups is provided in essence by the noise term, since
both speculators and producers contribute to $C$. In fact, the
correlation function can be written as
\begin{equation}
C(t,t') = n_s\alpha C_s(t,t')+n_p\alpha
\label{corr}
\end{equation}
where $n_s\alpha$ (resp. $n_p\alpha$) is the fraction of speculators
(resp. producers), and $C_s$ (resp. $C_p$) denote the correlation
function of speculators (resp. producers).  For the response function
we have, similarly,
\begin{eqnarray}
G(t,t') & = & n_s\alpha G_s(t,t')+n_p\alpha G_p(t,t')\nonumber\\ & = &
n_s\alpha G_s(t,t')
\label{resp}
\end{eqnarray}
producers being `frozen' at $n=1$ and thus insensitive to small
perturbations.

Assuming time-translation invariance,
\begin{gather}\label{tti}
\lim_{t\to\infty}C(t+\tau,t)=C(\tau)\\
\lim_{t\to\infty}G(t+\tau,t)=G(\tau)
\end{gather}
finite integrated response,
\begin{equation}\label{fir}
\lim_{t\to\infty}\sum_{t'\leq t}G(t,t')<\infty
\end{equation}
and weak long-term memory,
\begin{equation}\label{wltm}
\lim_{t\to\infty}G(t,t')=0~~~~~~~\forall t'\text{~finite}
\end{equation}
ergodic steady states of (\ref{eff}) can be characterized in terms of
a couple of order parameters, namely the persistent autocorrelation
\begin{equation}\label{c1}
c=\lim_{\tau\to\infty}\frac{1}{\tau}\sum_{t<\tau}C(\tau)
\end{equation}
and the integrated response (or susceptibility)
\begin{equation}\label{chi}
\chi=\lim_{\tau\to\infty}\sum_{t<\tau}G(\tau)
\end{equation}
From (\ref{corr}) and (\ref{resp}) we get
\begin{gather}\label{aa}
c=n_s\alpha c_s+n_p\alpha\\ \chi=n_s \alpha\chi_s\label{ab}
\end{gather}
where $c_s$ and $\chi_s$ are the persistent autocorrelation and
susceptibility of speculators. For $\lambda=0$ one can formulate,
inspired by computer experiments, a simple Ansatz for the dynamics of
scores that allows to calculate these quantities exactly as functions
of $c$ and $\chi$, so that from (\ref{aa}) and (\ref{ab}) one may
retrieve the values of the persistent order parameters for any $n_s$
and $n_p$. In this case, ergodicity breaks down as $\chi$ diverges at
certain critical values of the parameters, thus violating
(\ref{fir}). This analysis, including the dynamical phase transition,
reproduces the phenomenology of the batch-GCMG without memory
remarkably well, at least in the ergodic regime.

The score $y(t)$ of speculators either grows linearly with time as
$y(t)\simeq v~t$ (in which case the agent is `frozen' at inactivity
with $n(t)=0$ if $v<0$ or activity with $n(t)=1$ if $v>0$), or keeps
oscillating about $y(t)=0$ (in which case the agent is `fickle')
\cite{CM99}. In order to distinguish between the two situations, we
introduce the variable $\widetilde{y}(t)=y(t)/t$. Using this, we can
sum (\ref{eff}) over time to obtain
\begin{multline}
\widetilde{y}(t+1)-\frac{1}{t}y(1)=-\frac{\alpha}{t}
\sum_{t',t''}(I+G)^{-1}(t',t'')n(t'')\\-\alpha\epsilon+\frac{\sqrt{
\alpha}}{t}\sum_{t'}z(t')
\end{multline}
where $n(t)=\Theta[\widetilde{y}(t)]$. In the limit $t\to\infty$, the
above leads, via (\ref{tti}--\ref{wltm}), to a simple equation for the
quantity $\widetilde{y}=\lim_{t\to\infty}\widetilde{y}(t)$:
\begin{equation}\label{wty}
\widetilde{y}=-\frac{\alpha n}{1+\chi}-\alpha\epsilon+\sqrt{\alpha}z
\end{equation}
where $\chi$ is given by (\ref{chi}),
$n=\lim_{\tau\to\infty}(1/\tau)\sum_{t\leq\tau}n(t)$, and
$z=\lim_{\tau\to\infty}(1/\tau)\sum_{t\leq\tau}z(t)$ is a zero-average
Gaussian rv with variance
\begin{equation}
\avg{z^2}=\frac{1}{\tau\tau'}\sum_{t\leq\tau,t'\leq\tau'}
\avg{z(t)z(t')}=\frac{c}{(1+\chi)^2}
\end{equation}

Defining $\gamma=\sqrt{\alpha}/(1+\chi)$, we can proceed as usual by
separating the frozen speculators from the fickle ones. We have the
following situation: for $\widetilde{y}>0$ the effective speculator is
always active ($n=1$) and $z>\gamma+\sqrt{\alpha}\epsilon$; for
$\widetilde{y}<0$ the effective speculator is always inactive ($n=0$)
and $z<\sqrt{\alpha}\epsilon$; for $\widetilde{y}=0$ the effective
speculator is fickle, $n=\frac{z-\sqrt{\alpha}\epsilon}{\gamma}$ and
$\sqrt{\alpha}\epsilon<z<\gamma+\sqrt{\alpha}\epsilon$. So we have
\begin{multline}\label{cc}
c_s=\avg{\Theta(z-\gamma-\sqrt{\alpha}\epsilon)}\\
+\avg{\Theta(z-\sqrt{\alpha}\epsilon)\Theta(\gamma+\sqrt{\alpha}\epsilon-z)
\l(\frac{z-\sqrt{\alpha}\epsilon}{\gamma}\r)^2}
\end{multline}
with brackets denoting an average over $z$.  For $\chi_s$, we use the
fact that the noise $z(t)$ formally acts like an external source in
(\ref{wty}), so that
$\chi_s=(1/\sqrt{\alpha})\avg{\frac{\partial n}{\partial z}}$. This
gives
\begin{equation}\label{chichi}
\chi_s=
\frac{1}{\gamma\sqrt{\alpha}}\avg{\Theta(z-\sqrt{\alpha}\epsilon)
\Theta(\gamma+\sqrt{\alpha}\epsilon-z)}
\end{equation}
One can also calculate the average activity level of speculators as
\begin{multline}
\phi\equiv\avg{n}=\avg{\Theta(z-\gamma-\sqrt{\alpha}\epsilon)}+\\
\avg{\Theta(z-\sqrt{\alpha}\epsilon)\Theta(\gamma+\sqrt{\alpha}\epsilon-z)
(\frac{z-\sqrt{\alpha}\epsilon}{\gamma})}
\end{multline}
from which the number of active speculators per pattern follows as
$n_{\text{act}}=n_s\phi$, and the fraction of frozen speculators as
\begin{equation}
f=\avg{\Theta(z-\gamma-\sqrt{\alpha}\epsilon)}+
\avg{\Theta(\alpha\epsilon-z)}=f_1+f_0
\end{equation}
where $f_1$ (resp. $f_0$) stands for the fraction of always active
(resp. inactive) speculators. Inserting (\ref{cc}) and (\ref{chichi})
into (\ref{aa}) and (\ref{ab}) one obtains two equations that can be
solved self-consistently for $c$ and $\chi$. These equations can be
analyzed for any $\epsilon$. For the sake of simplicity, we focus on
the case $\epsilon=0$, in which the averages over $z$ take a
particularly simple form. We have
\begin{multline}
c=n_p\alpha+n_s\alpha\Bigg[\frac{1}{2}\l(1-
\erf\sqrt{\frac{\alpha}{2c}}\r)\\+\frac{c}{2\alpha}
\erf\sqrt{\frac{\alpha}{2c}}-e^{-\frac{\alpha}{2c}}
\sqrt{\frac{c}{2\pi\alpha}}\Bigg]\label{abb}
\end{multline}
\begin{equation}
\frac{\chi}{1+\chi}=\frac{n_s}{2}~\erf\sqrt{\frac{\alpha}{2c}}
\label{abc}
\end{equation}
whereas $\phi$ reads
\begin{equation}
\phi=\frac{1}{2}\l(1-\erf\sqrt{\frac{\alpha}{2c}}\r)+
\sqrt{\frac{c}{2\pi\alpha}}(1-e^{-\frac{\alpha}{2c}})
\end{equation}
As a quick consistency check, one can see, starting from (\ref{abc})
and with minor manipulations, that a divergence of the susceptibility
(i.e. the violation of (\ref{fir}) with consequent ergodicity
breaking) for $n_p=1$ occurs at $n_s=n_s^*=2/\erf(\xi^*)$ where
$\xi^*$ is the solution of the transcendental equation
$e^{-\xi^2}=\xi\sqrt{\pi}$. The result, $n_s^*=4.14542\ldots$, is in
full agreement with the replica results of both the on-line and the
batch-model.

\begin{figure}[t]
\includegraphics*[width=8.75cm]{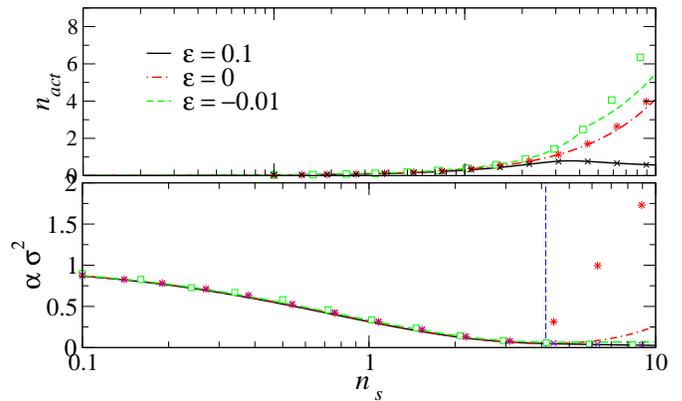}
\caption{\label{uno}GCMG with $\lambda=0$: number of active
speculators per pattern $n_{\text{act}}=n_s\phi$ (top) and volatility
(bottom) at $n_p=1$ as a function of $n_s$ for different values of
$\epsilon$. The dashed vertical line marks the position of the
critical point $n_s^*$. Markers denote results from computer
experiments.}
\end{figure}

In order to compare with the computer experiments discussed above, we
analyze the fraction of active speculators and the volatility
$\sigma^2$ obtained from (\ref{abb}--\ref{abc}) at $n_p=1$. As said
above, the former is just $n_{\text{act}}=n_s\phi$. As for the latter,
it is formally given by
\begin{equation}\label{esse2}
\sigma^2=\lim_{t\to\infty}\avg{z(t)z(t)}/\alpha
\end{equation}
It is possible to derive an approximate expression for the above limit
in terms of persistent order parameters assuming that the retarded
self-interaction of fickle speculators is negligible, that is, by
neglecting the agent's auto-correlation. This leads to
\cite{CoolenBatch}
\begin{equation}
\alpha\sigma^2=\frac{n_p+n_s f_1}{(1+\chi)^2}
+\frac{1}{4}n_s(1-f).
\end{equation}
Comparing with the replica result, one sees that $\sigma^2$ comprises
a $H_0$ term, as expected from the relationship (\ref{25}). However,
the other terms differ, as in the standard MG.  In Fig. \ref{uno} we
report the behavior of $\phi$ and $\sigma^2$ obtained from
(\ref{abb}--\ref{abc}) at $n_p=1$.
\begin{figure}[b]
\includegraphics*[width=8.75cm]{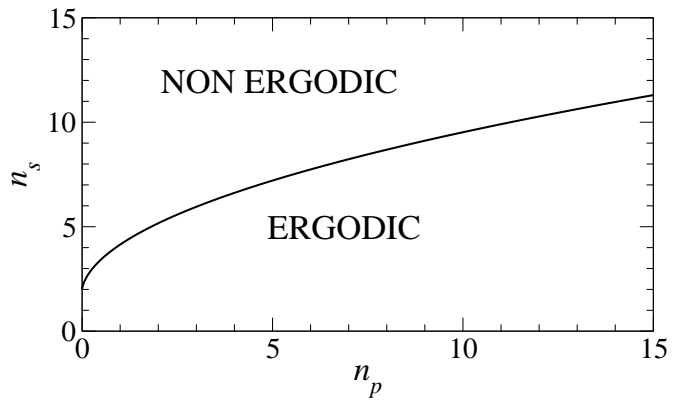}
\caption{\label{due}GCMG with $\lambda=0$: critical line $n_s^*$ vs
$n_p$, where $\chi$ diverges and (\ref{fir}) is violated at
$\epsilon=0$. The dynamics is ergodic for $n<n_s^*$.}
\end{figure}
Dynamical results for the batch model are in excellent agreement with
the simulations of the batch-GCMG and reproduce qualitatively the
phenomenology of the on-line-GCMG.  For the sake of completeness, we
also report (see Fig. \ref{due}), for $\epsilon=0$, the critical line
$n_s^*(n_p)$ where $\chi$ diverges and ergodicity breaks down. After
some algebra, it turns out to be given by $n_s^*(n_p)=2/\erf(\xi^*)$,
where $\xi^*\equiv\xi^*(n_p)$ is the solution of
\begin{equation}
\frac{e^{-\xi^2}}{\xi\sqrt{\pi}}=(n_p-1)\erf(\xi)+1
\end{equation}

\section{Stationary state with $\lambda>0$}
\begin{figure}
\centerline{\includegraphics[width=8.5cm]{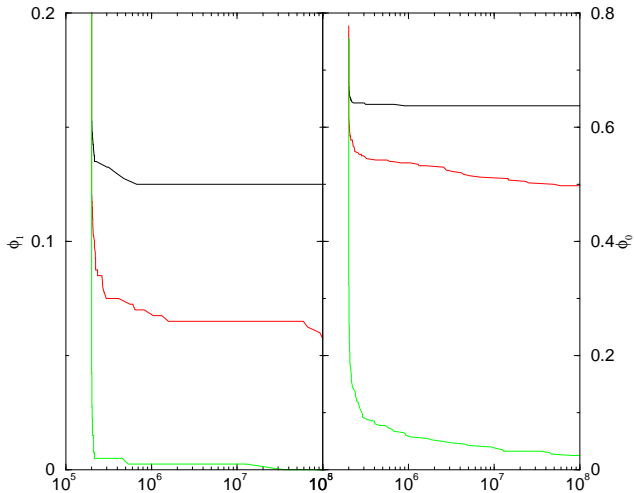}}
\caption{$\phi_1$ and $\phi_0$ as a function of time for games with
$\lambda=0$ (black lines), $0.01$ (red lines) and $0.1$ (green
lines). $P=100$, $n_s=4$, $\eps=0.05$}
\label{nomorephi1}
\end{figure}
One may expect that the stationary state of Minority Games with
$\lambda>0$ depends smoothly on $\lambda$ in ergodic regions, and
indeed some important quantities such as the fluctuations and $H$ do
behave this way. More surprising is the vanishing of frozen agents:
Fig. \ref{nomorephi1} reports that both $\phi_0$ and $\phi_1$ seem to
cancel for any positive $\lambda$, although this may not appear for
finite-time simulations at small values of $\lambda$; the measure was
done from $t=1000$ and counts the fraction of agents that are never
out of the market and in the market, respectively.

\subsection{On-line GCMG: static approach}

In many versions of the minority game with infinite score memory
($\lambda=0$) and na\"\i ve agents, a phase transition takes
place. $H$ behaves like a physical order parameter which is minimized
by the dynamics. The latter is ergodic as long as $H>0$. However, when
$H=0$, the stationary state is note unique, and the dynamics becomes
non-ergodic: the stationary state is selected by the initial score
valuation $y_i(0)$; this happens in particular in the original MG and
in the present GCMG with $\eps=0$ and $\lambda=0$. On the other hand,
it is obvious that if $y_i(0)$ is gradually forgotten, the stationary
state cannot depend anymore on $U_i(0)$. This is precisely what the
introduction of the finite score memory does: when $\lambda>0$, the
dynamics is ergodic, and accordingly the stationary state is
unique. This would be compatible with a minimized quantity that is not
cancelled anymore by the dynamics; intuitively, this means that a new
term is added to $H_\eps$. Unfortunately, finding such function turns
out to be impossible in this case, which rules out the use of replica-based approach.

\subsection{Batch GCMG: dynamical approach}

Let us now turn to the effective process (\ref{eff}) with finite score
memory. As discussed above, in this case scores do not diverge with
time, i.e. $\lim_{t\to\infty}y(t)<\infty$, and it is no longer
possible to separate frozen agents from fickle ones by the use of the
quantity $\widetilde{y}=\lim_{t\to\infty}y(t)/t$. Indeed, proceeding
as done for the case $\lambda=0$, one obtains, in place of
(\ref{wty}), the condition
\begin{equation}
\lambda\lim_{t\to\infty}\frac{1}{t}\sum_{t'\leq t}y(t')=-
\frac{\alpha\phi}{1+\chi}-\alpha\epsilon+\sqrt{\alpha}z
\end{equation}
where the term on the l.h.s. is finite. In addition, the fact that the
fraction of frozen players undergoes an extremely slow dynamics
ultimately suggesting that all agents are fickle in the stationary
state indicates the necessity of a different analytical approach when
$\lambda\neq 0$. Unfortunately, have been unable to capture the
peculiarities of the score dynamics with a simple Ansatz.  

\begin{comment}
\subsubsection{Idea from Damien}

The idea is to compute the distribution $P(\phi)$ starting from the
fact that a given realisation of $z(t)$ has an average of
$\ovl{z}=\lim_{T\to\infty}1/T\sum_{t\le T}z(t)$ whose distribution is
known to be Gaussian, zero mean and fluctuations $c/(1+\chi)^2$,
leading to $\lambda
y^*=\frac{\alpha\phi}{1+\chi}-\alpha\epsilon+\sqrt{\alpha}\ovl{z}$
(checked numerically). During its realisation, $z$ itself fluctuates
according to
$\zeta^2(\ovl{z}=\lim_{T\to\infty}1/T[z(t)-\ovl{z}]^2=\sigma^2-\ovl{z}^2$. We
are not interested in the memory of $z(t)$, but only in computing the
fraction of times $y>0$, which is obtained naively by assuming no
memory at all,
i.e. $P(y>0|\ovl{z})=\erf(y^*/\sqrt{2\zeta^2(\ovl{z})})/2=P(\phi|\ovl{z})$,
which is a self-consistent equation, since $\phi$ appears in
$y^*$. Hence
$P(\phi)=\int_{-\infty}^{\infty}\dd\ovl{z}P(\phi|\ovl{z})P(\ovl{z})$
\end{comment}

Some general hints can be obtained by calculating explicitly the first
time step of (\ref{eff}). For simplicity, we henceforth adopt the
shorthand $\avg{z(t)z(t')}=L(t,t')$ for (\ref{pop}). Furthermore, we
assume an initial condition $y(0)$ for (\ref{eff}) ensuring that all
speculators are active at time $0$, so that $C(0,0)=1$ and (by
causality) $G(0,0)=0$ and $L(0,0)=1$. The transition probability to
pass from $y(0)$ to $y(1)$ is given by
\begin{equation}
p[y(1)|y(0)]=\frac{1}{\sqrt{\alpha}} e^{-\frac{1}{2\alpha}[
y(1)-(1-\lambda)y(0)-\Theta(0)+\alpha n(0)+\alpha\epsilon]^2}
\end{equation}
As a consequence, we have
\begin{multline}
C(1,0)=\int dy(1)dy(0)p[y(1)|y(0)]p(y(0)) n(1) n(0)\\
=\frac{1}{2}\l[1-\erf\l(\frac{\alpha n(0)+\alpha\epsilon-
h(0)-(1-\lambda)y(0)}{\sqrt{2\alpha}}\r)\r]\\
=\frac{1}{2}\l[1-\erf\l(\frac{\alpha+\alpha\epsilon-(1-\lambda)
y(0)}{\sqrt{2\alpha}}\r)\r]
\end{multline}
where we set $h(0)=0$ and used the fact that $ n(0)=1$, and
\begin{multline}
G(1,0)=\frac{\partial}{\partial h(0)}\int
dy(1)dy(0)p[y(1)|y(0)]p(y(0)) n(1)\\ =\frac{1}{\sqrt{2\pi\alpha}}
\exp\l\{-\frac{1}{2\alpha}[\alpha+\alpha\epsilon-(1-\lambda)y(0)]^2\r\}
\end{multline}
From these, we can evaluate (recalling that in the steady state
(\ref{esse2}) holds) the time-dependent volatilities
\begin{multline}
L(1,0)=\sum_{t,t'}(I+G)^{-1}(1,t)C(t,t')(I+G^T)^{-1}(t',0)\\
=-G(1,0)+C(1,0)
\end{multline}
and
\begin{multline}
L(1,1)=\sum_{t,t'}(I+G)^{-1}(1,t)C(t,t')(I+G^T)^{-1}(t',1)\\
=G(1,0)^2-2C(1,0)G(1,0)+C(1,0)
\end{multline}
From the above formulae we see that if the initial condition is large,
in particular for $y(0)\gg\sqrt{\alpha}$, we have
\begin{gather}
\lim_{n_s\to\infty}C(1,0)=1\qquad\lim_{n_s\to\infty}G(1,0)=0\\
\lim_{n_s\to\infty}L(1,0)=1\qquad\lim_{n_s\to\infty}L(1,1)=1
\end{gather}
for $0<\lambda<1$, and
\begin{gather}
\lim_{n_s\to\infty}C(1,0)=0\qquad\lim_{n_s\to\infty}G(1,0)=0\\
\lim_{n_s\to\infty}L(1,0)=0\qquad\lim_{n_s\to\infty}L(1,1)=1
\end{gather}
for $\lambda>1$, provided $n_p$ is finite. The latter limits indicate
that, as is to be expected, for $\lambda>1$ the agent de-activates
immediately after the first time step and starts being active and
inactive alternatively. The former limits imply instead that the
effective agent continues to play. In particular, as long as he's
playing, he's insensitive to small perturbations, so that
\begin{equation}
y(t)\simeq (1-\lambda)y(0)+t\sqrt{\alpha}z(0)
\end{equation}
Hence one sees that the de-freezing occurs for times of the order of
$(1-\lambda)/\sqrt{\alpha}$.

\begin{figure}
\includegraphics*[width=8.5cm]{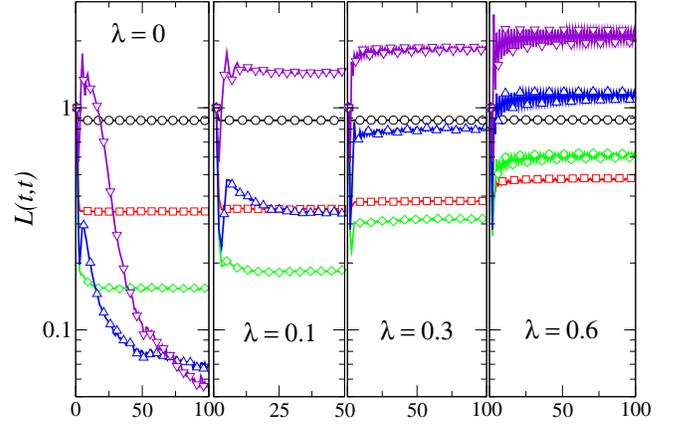}
\caption{\label{eo-neps}GCMG with different values of $\lambda$:
numerical solution of the effective dynamics for $M=250000$ copies
with $\epsilon=-0.1$. Markers correspond to the volatility as a
function of time for $n_s=0.1$ (circles)$,1$ (squares)$,2$
(diamonds)$,4$ (up triangles)$,8$ (down triangles). Only a few markers
are shown for simplicity.}
\end{figure}

In order to obtain a deeper insight on the stationary states of
(\ref{eff}), we now turn to a different approach, namely the
Eissfeller-Opper method.  In a nutshell, the idea is to simulate many
copies of the effective dynamics and calculate relevant physical
observables as averages over the whole population. The core of the
procedure lies in the possibility of evaluating the response function
without actually adding an external probing field. In fact, for
$G(t,t')$, which is formally given by $\avg{\frac{\partial
n(t)}{\partial h(t')}}$, one can again resort to the noise and write
\begin{eqnarray}
G(t,t') & = & \frac{1}{\sqrt{\alpha}}\avg{\frac{\partial
 n(t)}{\partial z(t')}}\nonumber\\ & = & \frac{1}{\sqrt{\alpha}} \int
 \frac{\partial n(t)}{\partial z(t')} P(z) Dz
\end{eqnarray}
Now the noise distribution $P(z)$ is
\begin{equation}
P(z)\sim \exp\l[-\frac{1}{2}\sum_{t,t'}z(t)L^{-1}(t,t')z(t')\r]
\end{equation}
so that, after an integration by parts, one gets
\begin{eqnarray}
G(t,t') & = &
 \sum_{t''=0}^{t-1}\avg{ n(t)z(t'')}L^{-1}(t'',t')\nonumber\\ &
 \equiv & \sum_{t''=0}^{t-1} K(t,t'')L^{-1}(t'',t')
\end{eqnarray}
The matrix $K$, as well as the correlation function $C$, can be
evaluated by an average over the copies (say, $M$) of the effective
dynamics:
\begin{gather}
K(t,t')=\frac{1}{M}\sum_{\ell=1}^M n_\ell(t)z_\ell(t')\\
C(t,t')=\frac{1}{M}\sum_{\ell=1}^M n_\ell(t) n_\ell(t')
\end{gather}
and the only remaining problem is that of generating a noise $z(t)$
having the desired statistical properties. This can be done by
properly summing and re-scaling unit Gaussian variables. We focused
again on the time-evolution of the quantity $L(t,t)$, whose limit
$t\to\infty$ is linked to the volatility (see (\ref{esse2})). Results
for different $\lambda$'s and $n_s$ at $n_p=1$ are shown in
Fig. (\ref{eo-neps}). One sees that finite memory generally increases
fluctuations. The effect is more pronounced for large values of $n_s$.

\section{Conclusions}

To summarzie, we have studied the effects induced by a finite score
memory in MG-based market models. Our main result is that finite score
memory does not destroy market-like phenomenology in grand canonical
minority games, and remedies two embarassing problems of minority
games with a fixed set of strategies. From this point of view,
implementing finite scores is a useful extension of GCMG, but cannot be understood fully by 
current analytical methods.

\appendix
\section{Original MG with $\lambda>0$.}
\label{appendix:MG}

The definition of the original MG is similar to that of the GCMG,
except that all the agents have at least two strategies, and one score
per strategy. Assuming that they have only two strategies, denoted
$a_{i,1}$ and $a_{i,2}$ for agent $i$, each having their respective
score $y_{i,1}$ and $y_{i,2}$. The latter evolve according to
\be\label{MGon-line}
y_{i,s}(t+1)=y_i(t)(1-\lambda/P)-a_{i,s}^{\mu(t)}A(t) \ee for the
on-line MG, and \be\label{MGbatch}
y_{i,s}(t+1)=y_i(t)(1-\lambda)-a_{i,s}^{\mu(t)}A(t) \ee for the batch
MG. At each time step, the agents use their strategy that has the
highest score. The original MG was defined with $\lambda=0$, and the
usual control parameter is $\alpha=P/N$. There is a phase transition
at $\alpha_c=0.3374\ldots$ that separates an asymmetric, information
rich phase with $H>0$ ($\alpha>\alpha_c$) from a symmetric phase with
$H=0$ ($\alpha<\alpha_c$). In the latter, the stationary state depends
on the asymmetry
\[
U_0=U_{i,1}(0)-U_{i,2}(0)
\] 
of the initial conditions. This dependence is due to the fact that,
if $\lambda=0$,, all market fluctuations since time $t=0$ are
remembered and contribute with the same weight to $U_{i,s}(t)$,
irrespective of how far in the past they took place.

\begin{figure}
\centerline{\includegraphics[width=0.4\textwidth]{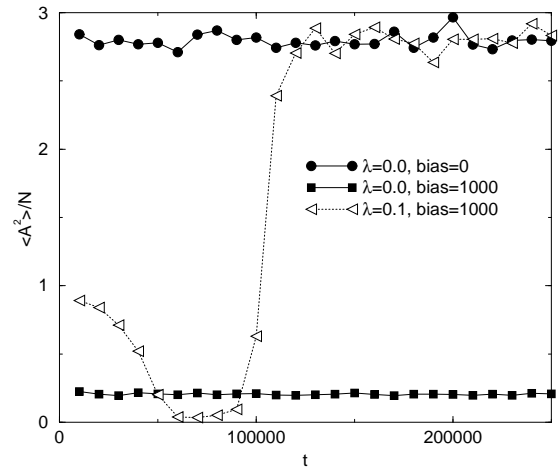}}
\caption{The same realization of the standard minority game with
  unbiased initial strategy scores and $\lambda=0$ (circles), biased
  initial scores and $\lambda=0$ (squares), biased initial scores and
  $\lambda=0.1$ (triangles). ($P=100$, $\alpha=0.1$)}
\label{s2t}
\end{figure}

Fig. \ref{s2t} shows that a time dependent $\sigma^2=\avg{A^2}_t$
(here, $\avg{\ldots}_t$ stands for an average over a long but finite
time interval around $t$). converges, for long times, to a value which
is independent of initial conditions $U_0$. This implies that the
initial asymmetry $U_0$ has no influence any more on the stationary
state. 

\begin{figure}
\centerline{\includegraphics[width=0.4\textwidth]{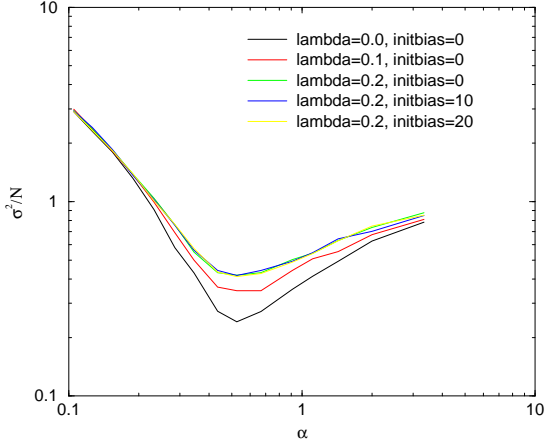}}
\caption{Average volatility $\sigma^2/N=\avg{A^2}/N$ versus
  $\alpha=P/N$ of the standard MG with unbiased initial scores and
  $\lambda=0$ (circles), $\lambda=0.1$ (squares), $\lambda=0.2$
  (diamonds), with biased initial scores and $\lambda=0.2$
  (triangles).}
\label{s2alpha}
\end{figure}

\begin{figure}
\centerline{\includegraphics[width=0.4\textwidth]{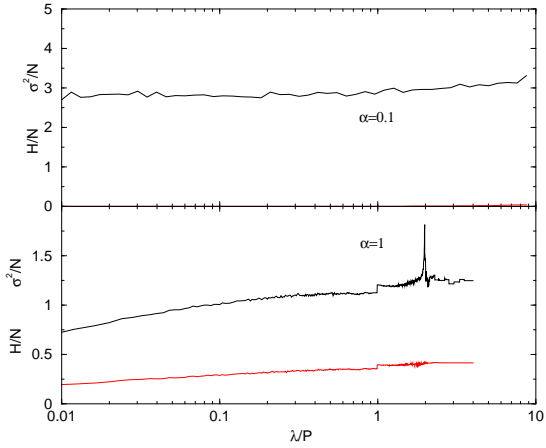}}
\caption{$\sigma^2/N$ (black lines) and $H/N$ (red lines) for a realization of the standard Minority Game with increasing $\lambda$}
\label{s2Hlambda}
\end{figure}

This is confirmed by Fig. \ref{s2alpha}, where we plot
$\sigma^2/N$ as a function of $\alpha$ for various $U_0$ and
$\lambda$. For a fixed $\alpha$ and for a given realization of the
game, $\sigma^2$ and $H$ are increasing functions of $\lambda$,
although in the symmetric phase ($\alpha<\alpha_c=0.3374\dots$)
$\sigma^2$ varies very slowly with $\lambda$ (see
Fig. \ref{s2Hlambda}). While only small values of $\lambda$ make
physical sense, for completeness the same figure reports also the
strange effects of very large $\lambda$. From \req{MGon-line},
$\lambda=P$ cancels the contribution of $U_i(t)$ in $U_i(t+1)$; larger
$\lambda$ implies that this contribution is of opposite sign to the
usual MG.

%\begin{references}
%\bibitem{bs}

%\vfill\eject
%\onecolumn

%\begin{figure}
%\centerline{\includegraphics[width=0.4\textwidth]{absfig1.eps}
%\caption{}
%\label{fig1}
%\end{figure}

\end{document}